\documentclass[12pt, amsfonts]{article}
\begin{document}
\def\p {{\partial}}
\def\n {{\nu}}
\def\m {{\mu}}
\def\a {{\alpha}}
\def\bt {{\beta}}
\def\f {{\phi}}
\def\th {{\theta}}
\def\g {{\gamma}}
\def\eps {{\epsilon}}
\def\e {{\psi}}
\def\la {{\lambda}}
\def\na {{\nabla}}
\def\bn {\begin{eqnarray}}
\def\en {\end{eqnarray}}
\title{Canonical path integral quantization of the finite dimensional
systems with constraints\footnote{ E-Mail
$sami_{-}muslih$@hotmail.com}} \maketitle
\begin{center}
\author{S.I. MUSLIH\\Dept. of Physics\\ Al-Azhar university\\
Gaza, Palestine}
\end{center}

\begin{abstract}
The path integral formulation of constrained systems leads to
obtain the equations of motion as total differential equations in
many variables.  If these equations are integrable then one can
construct a valid and a canonical phase space coordinates. The
path integral is obtained as an integration over the canonical
phase space coordinates. This approach is applied to obtain the
path integral for three singular systems and it shown that in our
formulation there is no need to distinguish between  first and
second- class constraints, no need for fixing any gauge, as will
as no need to enlarge the phase space.
\end{abstract}
\newpage
\section{Introduction}

The canonical formulation [1-3] gives the set of Hamilton-Jacobi
partial differential equations (HJPDE) as \bn
&&H^{'}_{\a}(t_{\bt}, q_a, \frac{\p S}{\p q_a},\frac{\p S}{\p
t_a}) =0,\nonumber\\&&\a, \bt=0,n-r+1,...,n, a=1,...,n-r,\en where
\begin{equation}
H^{'}_{\a}=H_{\a}(t_{\bt}, q_a, p_a) + p_{\a},
\end{equation}
and $H_{0}$ is defined as
\bn
 &&H_{0}= p_{a}w_{a}+ p_{\m} \dot{q_{\m}}|_{p_{\n}=-H_{\n}}-
L(t, q_i, \dot{q_{\n}},
\dot{q_{a}}=w_a),\nonumber\\&&\m,~\n=n-r+1,...,n. \en

The equations of motion are obtained as total differential
equations in many variables as follows:

\bn
 dq_a=&&\frac{\p H^{'}_{\a}}{\p p_a}dt_{\a};\\
 dp_a=&& -\frac{\p H^{'}_{\a}}{\p q_a}dt_{\a};\\
dp_{\bt}=&& -\frac{\p H^{'}_{\a}}{\p t_{\bt}}dt_{\a};\\
 dZ=&&(-H_{\a}+ p_a \frac{\p
H^{'}_{\a}}{\p p_a})dt_{\a};\\
&&\a, \bt=0,n-r+1,...,n, a=1,...,n-r\nonumber \en where
$Z=S(t_{\a};q_a)$. The set of equations (4-7) is integrable [3] if
\bn
dH^{'}_{0}=&&0,\\
dH^{'}_{\m}=&&0,  \m=n-p+1,...,n, \en or in equivalent form
\begin{equation}
[H^{'}_{\a},\;\;H^{'}_{\bt}]=0\;\;\forall\; \a,\bt.
\end{equation}
Equations of motion reveal the fact that the Hamiltonians
$H^{'}_{\a}$ are considered as the infinitesimal generators of
canonical transformations given by parameters $t_{\a}$ and the set
of canonical phase-space coordinates $q_a$ and $p_a$ is obtained
as functions of $t_{\a}$, besides the canonical action integral is
obtained in terms of the canonical coordinates. In this case, the
path integral representation may be written as [4-7]

\bn
D({q'}_a,{t'}_{\a};q_a,t_{\a})=&&\int_{q_a}^{{q'}_a}~Dq^{a}~Dp^{a}\times
\nonumber\\&&\exp i \{\int_{t_{\a}}^{{t'}_{\a}}[-H_{\a}+
p_a\frac{\p H^{'}_{\a}}{\p
p_a}]dt_{\a}\},\nonumber\\&&a=1,...,n-r, \a=0,n-r+1,...,n. \en

Now we will study the path integral quantization of the finite
dimensional systems considering systems with first and second
class constraints .
\section{Examples}
\subsection{A system with first class constraints}

As a first example let us consider the following Lagrangian [1]
\begin{equation}
L= \frac{1}{2}a_{ij}\dot{q_{i}}\dot{q_{j}}+ b\dot{q_{2}} -
c,\;\;i,j=1, 2, 3.
\end{equation}
The generalized momenta read as
\begin{equation}
p_{1}= a_{1} \dot{q_{1}},\;p_{2}=
a_{2}(\dot{q_{3}}-\dot{q_{2}})+b,\;p_{3}=
a_{2}(\dot{q_{3}}-\dot{q_{2}}).
\end{equation}
The canonical method [1-3] leads us to obtain the set of the
Hamilton- Jacobi partial differential equations as follows: \bn
&&{H'}_{0}= p_{0} + \frac{1}{2}(\frac{p_{1}^{2}}{a_{1}} -
\frac{p_{3}^{2}}{a_{2}}) + c=0,\\
&&{H'}_{2}= p_{2} + p_{3} -b =0, \en where $a, b$ and $c$ are
constants.

The equations of motion are obtained as total differential
equations in many variables as follows: \bn&& dq_{1}=\frac{\p
{H'}_{0}}{\p \p_{1}}dt + \frac{\p {H'}_{2}}{\p
p_{1}}dq_{2}=p_{1}dt,\\
&&dq_{3}=\frac{\p {H'}_{0}}{\p p_{3}}dt + \frac{\p {H'}_{2}}{\p
p_{3}}dq_{2}= - \frac{p_{1}}{a_{2}}dt +dq_{2},\\
&&dp_{1}=-\frac{\p {H'}_{0}}{\p q_{1}}dt - \frac{\p {H'}_{2}}{\p
q_{1}}dq_{2}=0,\\
&&dp_{2}=-\frac{\p {H'}_{0}}{\p q_{2}}dt - \frac{\p {H'}_{2}}{\p
q_{2}}dq_{2}=0,\\
&&dp_{3}=-\frac{\p {H'}_{0}}{\p q_{3}}dt - \frac{\p {H'}_{2}}{\p
q_{2}}dq_{2}=0,\\
&&dp_{0}=-\frac{\p {H'}_{0}}{\p t}dt - \frac{\p {H'}_{2}}{\p t
}dq_{2}=0.
\en

From the integrability conditions (8,9) since the total variations
$d{H'}_{0}=0, d{H'}_{2}=0$ are satisfied identically, the set of
equations are integrable and the canonical phase space
coordinates are obtained in terms of parameters $t$ and $q_{2}$
as follows: \bn&& q_{1}\equiv q_{1}(t, q_{2}),\;\;p_{1}\equiv
p_{1}(t, q_{2}),\\
&& q_{3}\equiv q_{3}(t, q_{2}),\;\;p_{3}\equiv p_{3}(t, q_{2}).
\en

Making use of (7) and (14, 15), the canonical action integral is
obtained as follows
\begin{equation}
z=\int\{ ( -c +
\frac{p_{1}^{2}}{2a_{1}}-\frac{p_{3}^{2}}{2a_{2}})dt + b dq_{2}\}.
\end {equation}

Now the path integral for the given system is obtained as an
integration over the canonical phase space coordinates $(q_{1},
p_{1}; q_{3}, p_{3})$ as follows: \bn D({q'}_1,{q'}_3, t',{q'}_{2}
; {q}_1,{q}_3, t,{q}_{2})&&=\int_{q_{1}, q_{3} }^{{q'}_{1},
{q'}_{3} }~Dq_{1}~Dq_{3}~Dp_{1}~Dp_{3}\times \nonumber\\&&\exp i
[\int_{t, q_{2}}^{t', {q'}_{2}} ( -c +
\frac{p_{1}^{2}}{2a_{1}}-\frac{p_{3}^{2}}{2a_{2}})dt + b dq_{2}].
\en

\subsection{A system with second class constraints}

Let us consider the singular Lagrangian
\begin{equation}
L= \frac{1}{2}{\dot q_{1}}^{2} - \frac{1}{4}({\dot q_{2}}^{2}
-2\dot {q_{2}}\dot{q_{3}}+ {\dot q_{3}}^{2}) +(q_{1} +
q_{3})\dot{q_{2}} -q_{1}- q_{2} -q_{3}^{2}.
\end{equation}

The generalized momenta read as
\begin{equation}
p_{1}= \dot{q_{1}},\;p_{2}=\frac{1}{2}(\dot{q_{3}}-\dot{q_{2}})+
q_{1} +q_{3},\;p_{3}= \frac{1}{2}(\dot{q_{2}}-\dot{q_{3}}).
\end{equation}
Since the rank of the Hessian matrix is two one of the momenta is
depending on the others. Thus we have \bn
&&\dot{q_{1}}=p_{1}=w_{1},\;\;\dot{q_{3}}= \dot{q_{2}}- 2p_{3},\\
&&p_{2} = - p_{3} + q_{1} +q_{3} = - H_{2}. \en The Hamiltonian
$H_{0}$ is defined as
\begin{equation}
H_{0} = - L(q_{1}, q_{2}, \dot {q_{2}}, w_{1}, w_{3}) + p_{1}w_{1}
+ p_{3}w_{3} + (-  p_{3} + q_{1} +q_{3})\dot{q_{2}},
\end{equation}
or
\begin{equation}
H_{0} = \frac{1}{2}(p_{1}^{2} - 2p_{3}^{2}) +q_{1} +q_{2} +
q_{3}^{2}.
\end{equation}
Thus the Hamiltonians $H{'}_{0}$ and ${H'}_{2}$ are obtained as
\bn &&H{'}_{0}= p_{0}+ \frac{1}{2}(p_{1}^{2} - 2p_{3}^{2}) +q_{1}
+q_{2} + q_{3}^{2}=0,\\
&& {H'}_{2}= p_{2} + p_{3} - q_{1} - q_{3}= 0. \en Making use of
(32) and (33) the equations of motion of this system are obtained
as \bn&& dq_{1}=\frac{\p {H'}_{0}}{\p \p_{1}}dt + \frac{\p
{H'}_{2}}{\p
p_{1}}dq_{2}=p_{1}dt,\\
&&dq_{3}=\frac{\p {H'}_{0}}{\p p_{3}}dt + \frac{\p {H'}_{2}}{\p
p_{3}}dq_{2}= - 2p_{3}dt +dq_{2},\\
&&dp_{1}=-\frac{\p {H'}_{0}}{\p q_{1}}dt - \frac{\p {H'}_{2}}{\p
q_{1}}dq_{2}=-dt + dq_{2},\\
&&dp_{2}=-\frac{\p {H'}_{0}}{\p q_{2}}dt - \frac{\p {H'}_{2}}{\p
q_{2}}dq_{2}=-dt ,\\
&&dp_{3}=-\frac{\p {H'}_{0}}{\p q_{3}}dt - \frac{\p {H'}_{2}}{\p
q_{2}}dq_{2}=-2q_{3}dt + dq_{2},\\
&&dp_{0}=-\frac{\p {H'}_{0}}{\p t}dt - \frac{\p {H'}_{2}}{\p t
}dq_{2}=0. \en

To check whether this system is integrable or not, let us
consider the variation of ${H'}_{0}$, where
\begin{equation}
d{H'}_{0}= dp_{0} +dq_{1} +dq_{2} + p_{1}dp_{1} +2q_{3}dq_{3}-
2p_{3}dp_{3}.
\end{equation}
Making use of eqs. of motion (34-39) one obtains
\begin{equation}
d{H'}_{0}= {H'}_{3}dq_{2},
\end{equation}
where
\begin{equation}
{H'}_{3}= 2p_{3}-2q_{3} -p_{1} -1.
\end{equation}
Since ${H'}_{3}$ is not identically zero, we consider it as a new
constraint. Thus for a valid theory, variation of ${H'}_{3}$
should be zero. Thus one gets
\begin{equation}
d{H'}_{3}= (1- 4q_{3} + 4p_{3})dt- dq_{2}=0,.
\end{equation}
which can be expressed as
\begin{equation}
\dot{q_{2}} -(1-4q_{3} +4p_{3})=0.
\end{equation}
Again considering the variation of () one obtains
\begin{equation}
\ddot{q_{2}}-8( p_{3} -q_{3})=0.
\end{equation}
Combining Eqs. (44) and (45), differential equation for $q_{2}$ is
determined as
\begin{equation}
\ddot{q_{2}}-2\dot{q_{2}}+2=0,
\end{equation}
which has the following solution
\begin{equation}
q_{2}= 2 A\exp (2t) + t + C,
\end{equation}
where $A$ and $C$ are arbitrary constants. Besides the variation
of (33) is
\begin{equation}
d{H'}_{2}= dp_{2} +dp_{3} - dq_{1} -dq_{3},
\end{equation}
or
\begin{equation}
d{H'}_{2}= {H'}_{3}dt.
\end{equation}

The set of equations (34-39) is integrable, hence the canonical
action integral is calculated as
\begin{equation}
z=\int[(-H_{0}+ p_{1}^{2}- 2p_{3}^{2})dt + (- H_{2} +
p_{3})dq_{2}],
\end{equation}
or
\begin{equation}
z=\int(-H_{0}+ p_{1}^{2}- 2p_{3}^{2} +q_{1}+q_{3})dt.
\end{equation}
Making use of (11) and (51), the path integral for this system is
obtained as \bn D({q'}_1,{q'}_3, t',{q'}_{2} ; {q}_1,{q}_3,
t,{q}_{2})&&=\int_{q_{1}, q_{3} }^{{q'}_{1}, {q'}_{3}
}~Dq_{1}~Dq_{3}~Dp_{1}~Dp_{3}\times \nonumber\\&&\exp i
[\int_{t}^{t'} ( - H_{0}+ p_{1}^{2}- 2p_{3}^{2} +q_{1}+q_{3})dt].
\en
\section{Path integral for canonically transformed systems}
In this section we will consider the path integral quantization
of singular systems after performing some canonical
transformations

Let us perform the canonical transformations \bn &&t_{\a}=(t,
q_{\m})\rightarrow T_{\a}=(t,Q_{\m}),\;p_{0}\rightarrow
p_{0},\;q_{a}\rightarrow Q_{a},\;p_{a}\rightarrow
P_{a},\nonumber\\
&&a=1,...,n-r,\;\;\; \a=0,n-r+1,...,n,\;\;\;\m=n-r+1,...,n. \en
In this case the Hamiltonians ${H'}_{\a}$ are transformed as
follows
\begin{equation}
{H'}_{\a}\rightarrow{K'}_{\a}= P_{\a}+ K_{\a}=0 .
\end{equation}
The transformations (53) and (54) lead us to obtain the path
integral representation for this system as \bn
D({Q'}_a,{T'}_{\a};Q_a,T_{\a})=&&\int_{Q_a}^{{Q'}_a}~DQ^{a}~DP^{a}\times
\nonumber\\&&\exp i \{\int_{T_{\a}}^{{T'}_{\a}}[-K_{\a}+
P_a\frac{\p K^{'}_{\a}}{\p
P_a}]dT_{\a}\},\nonumber\\&&a=1,...,n-r, \a=0,n-r+1,...,n. \en

The procedure described above will be demonstrated by the
following example.

Let us consider the Lagrangian on the three-dimensional
configuration space $R^{3}=(x,y,z)$ [8]:
\begin{equation}
L= \frac{1}{2r^{2}}(x\dot{x} + y\dot{y}+ z\dot{z})^{2} - V(x^{2}
+ y^{2}+ z^{2}),
\end{equation}
where $r^{2}=x^{2} +y^{2} + z^{2}$.

The canonical momenta are obtained as
\begin{equation}
p_{x}= \frac{x}{2r^{2}}(x\dot{x} + y\dot{y}+ z\dot{z}),\;p_{y}=
\frac{y}{2r^{2}}(x\dot{x} + y\dot{y}+ z\dot{z}),\;p_{z}=
\frac{z}{2r^{2}}(x\dot{x} + y\dot{y}+ z\dot{z}).
\end{equation}
Since the rank of the Hessian matrix is one, the canonical method
leads us to obtain the set of Hamilton-Jacobi partial
differential equations as follows \bn &&{H'}_{0}= p_{0}+  V(x^{2}
+ y^{2}+ z^{2})+ \frac{p_{x}^{2}}{2x^{2}}( x^{2} + y^{2}+ z^{2})=0,\\
&&{H'}_{1}= p_{y}-y\frac{p_{x}}{x},\;\;\;{H'}_{2}=
p_{z}-z\frac{p_{x}}{x}. \en

The equations of motion are obtained as as set of total
differential equations as follows \bn&&dx =\frac{p_{x}}{x^{2}}(
x^{2}
+ y^{2}+ z^{2})dt -\frac{y}{x}dy -\frac{z}{x}dz,\\
&&dp_{x}= -\frac{\p V}{\p x}dt + \frac{p_{x}^{2}}{x^{3}}( y^{2}+
z^{2})dt -\frac{z p_{x}}{x^{2}}dy
-\frac{y p_{x}}{x^{2}}dz,\\
&&dp_{y}= (-\frac{\p V}{\p y} - \frac{y p_{x}^{2}}{x^{2}})dt +
\frac{ p_{x}}{x}dy,\\
&&dp_{z}= (-\frac{\p V}{\p z} - \frac{z p_{x}^{2}}{x^{2}})dt +
\frac{ p_{x}}{x}dz,\\
&&dp_{0}=0 . \en

To have  a consistent theory one should consider the variations
of ${H'}_{0}, {H'}_{1}$ and ${H'}_{2}$. In fact, one can show that
the total variations for each one of them is identically zero.
Hence, this system is integrable and the canonical phase space
coordinates $(x, p_{x})$ are obtained in terms of parameters $(t,
y, z)$ and one can use the procedure described in section (1) to
obtain the path integral for this system as an integration over
the canonical phase space coordinates $x, p_{x}$.

Now let us perform the canonical transformations \bn&&
{y\rightarrow y},\;\;\;{z\rightarrow z},\;\;\;{x\rightarrow
\sqrt{R^{2}- y^{2}-z^{2}}},\\
&&p_{x}\rightarrow \sqrt{R^{2}- y^{2}-z^{2}}\frac
{P_{R}}{R},\;\;p_{y}\rightarrow P_{y} +\frac{y
P_{R}}{R},\;\;p_{z}\rightarrow P_{z} +\frac{z P_{R}}{R}. \en The
corresponding set of Hamilton - Jacobi partial differential
equations read as \bn&&{K'}_{0}= p_{0} + \frac{P_{R}^{2}}{2} +
V(R^{2})=0,\\
&&{K'}_{1}= P_{y}=0,\\
&&{K'}_{2}= P_{z}=0. \en This set leads us to obtain the total
differential equations
\begin{equation}
dR= P_{R}dt,\;\;dP_{R}= -\frac{\p V}{\p
R}dt,\;\;dP_{y}=0,\;\;\;dP_{z}=0\,\;\;dp_{0}=0.
\end{equation}

Integrability conditions require the total variations of
${K'}_{0}, {K'}_{1}$ and ${K'}_{2}$ vanish. In fact
\begin{equation}
d{K'}_{0}= d{K'}_{1}=d{K'}_{2}=0.
\end{equation}
The set set of equations (70) is integrable and the canonical
phase space coordinates $(R, P_{R})$ are obtained in terms of
parameters $(t, y, z)$. Besides the canonical action integral is
calculated as
\begin{equation}
dz = (\frac{P_{R}^{2}}{2} -V(R^{2}))dt.
\end{equation}

Making use of (55) and (72) the path integral representation for
this system is obtained as
\begin{equation}
D(R, y, z, t ; R', y', z', t')=\int_{R}^{R'}~DR~DP_{R}\times \exp
i \{\int_{t}^{t'}[\frac{P_{R}^{2}}{2} -V(R^{2})]dt\}.
\end{equation}

\section{ Conclusion}

The Path integral formulation of constrained systems is obtained
using the canonical path integral method introduced by Muslih
[4-7]. The staring point of this method is variational principle.
The Hamiltonian treatment of constrained systems leads to a set
of Hamilton -Jacobi partial differential equations, which leads
to obtain the equations of motion as total differential equations
in many variables. The equations are integrable if the
corresponding system of partial differential equations is a
Jacobi system. In this case one can construct a valid and
canonical phase space coordinates $q_{a}$ and $p_{a}$ in terms of
parameters $t_{\a}$ and the path integral is obtained directly as
an integration over the canonical phase space coordinates $q_{a}$
and $p_{a}$.

In the first example since this system is integrable, then the
canonical phase space coordinates $(q_{1}, q_{3}, p_{1}, p_{3})$
are obtained as an integration over these canonical phase space
coordinates directly without using any gauge fixing conditions
[9, 10]. The second example is integrable and the path integral
is obtained as an integration over the canonical phase space
coordinates $(q_{1}, q_{3}, p_{1}, p_{3})$. in the usual
formulation [11] one has to integrate over the extended phase
space
 $(q_{1},q_{2}, q_{3}, p_{1},p_{2}, p_{3})$ and after integration over the
 redundant variables $(q_{2}, p_{2})$ one can arrive at the result
 (52).

 The path integral for singular systems is obtained when a suitable
 canonical transformation is used. For the system (56), after
 performing the canonical transformations (53), we obtain the
 equations of motion as total differential equations which are
 integrable. In this case canonical phase space coordinates $(R,
 P_{R})$ are obtained in terms of parameters $(t, y, z)$ and the
 path integral (73) is obtained as an integration over the canonical
 phase space coordinates $(R, P_{R})$.

 As a conclusion it is obvious that the Muslih [4-7] method is a direct
 method to obtain the path integral for constrained systems as an
 integration over the canonical phase space coordinates $(q_{a},
 p_{a})$. It is obvious from the given examples that when applying this method
 their is no  need to distinguish between first and second class constraintes,
 no need to use gauge fixing conditions, no need to enlarge the
 phase space, as will as  no need to add auxiliary dynamical variables
 expanding the phase space beyond its original classical
 formulation, including no ghosts. All is needed the set of the Hamilton Jacobi partial differential
equations and the set of the equations of motion. Then one should
tests whether these equations are integrable or not. If the
integrability conditions are not satisfied identically, then the
total variation of them should be introduced as new constraints
of the theory. Repeating this procedure as many times as needed
one may obtain a set of conditions. The number of independent
parameters of the theory is determined directly, without imposing
any gauge fixing conditions by this set.

\end{document}